# Unusual Hole-doping-dependent Electronic Instability and Electron-Phonon Coupling in Infinite-layer Nickelates

Xuelei Sui[1#], Jianfeng Wang[2,1#], Xiang Ding[3], Ke-Jin Zhou[4], Liang Qiao[3], Haiqing Lin[1,5], and Bing Huang[1,5*]

[1]*Beijing Computational Science Research Center, Beijing 100193, China*
[2]*School of Physics, Beihang University, Beijing 100191, China*
[3]*School of Physics, University of Electronic Science and Technology of China, Chengdu 610054, China*
[4]*Diamond Light Source, Harwell Campus, Didcot OX11 0DE, United Kingdom*
[5]*Department of Physics, Beijing Normal University, Beijing 100875, China*

[#]These authors contributed equally to this work.
[*]Correspondence should be addressed to B.H. (bing.huang@csrc.ac.cn).

The interplay between superconductivity and charge density waves (CDWs) under hole doping in cuprates is one of the central phenomena in condensed matter physics. Recently, CDWs are also observed in CaCuO$_2$-analogous nickelates $R$NiO$_2$ ($R$ = La, Nd) but exhibit fundamentally different hole-doping-dependent behaviors compared to that in cuprates, raising a challenging question on its origin. In this article, we propose that electronic instability (EI) and moment-dependent electron-phonon coupling (MEPC), mainly contributed by Ni $3d_{x^2-y^2}$ and $R$ $5d_{z^2}$, respectively, may be the possible reasons for CDW formation in $R$NiO$_2$. Without hole doping, a strong Fermi surface nesting (FSN) induced by the unique feature of van Hove singularity (VHS) across Fermi level exists in $R$NiO$_2$ but not in CaCuO$_2$, and the unusual temperature-insensitive feature of EI and MEPC could result in rather high temperature CDWs in $R$NiO$_2$. Under hole doping, the reduced FSN of Ni $3d_{x^2-y^2}$ by the shift of VHS and decreased occupation of $R$ $5d_{z^2}$ largely weaken EI and MEPC in $R$NiO$_2$, respectively, suppressing the CDW formation. Our theory not only offers possible explanations to some puzzling experimental observations, but also establishes a unified understanding on the hole-doping-dependent EI and MEPC in nickelates and cuprates.

***Introduction.*** Despite over 30 years of intense effort, the understandings of the physical origin accounting for the existence of high-temperature superconductivity [1-3] and competing symmetry-breaking orders [4,5] in cuprates remain the top challenges in condensed matter physics. A possible route to solve the puzzles in cuprates is to find other non-Cu-based superconductors but with cuprate-analogous structures [6,7]. Recently, the long-awaited superconductivity is eventually realized in hole-doped infinite-layer nickelates, *e.g.*, CaCuO$_2$-like $R$NiO$_2$ ($R$ = La, Nd) [8,9], potentially bringing us to the new age of nickelates.

Besides of the superconductivity, charge density waves (CDWs), which are popular but competing or intertwining with superconductivity in cuprates [4,5,10], are also observed in LaNiO$_2$ [11] and NdNiO$_2$ [12,13]. In cuprates, the CDWs are usually absent in undoped systems but appear under certain hole doping concentrations ($n_h$) with a dome [5,14]. However, in nickelates, the CDWs appear in undoped LaNiO$_2$ [11] and NdNiO$_2$ [12,13] but becomes to monotonously weaken and eventually disappear without a dome, as Sr-doping-induced $n_h$ increases, accompanied by the shift of $q_{CDW}$ vector [11]. The possible mechanism for forming CDWs in $R$NiO$_2$, which is even under debate in cuprates [15-17], is largely unclear at this stage, as they might have multiple origins [18,19]. Since there is no clear evidence that hole doping can significantly change the electron correlation in $R$NiO$_2$ [20,21], we pay our attention to electronic instability (EI) and moment-dependent electron-phonon coupling (MEPC) in $R$NiO$_2$, as they are two of the major driving forces to form CDWs in solids [22,23]. In particular, we focus on a unified understanding of $n_h$-dependent behaviors of EI and MEPC in nickelates and cuprates, which may shed light on their significantly different $n_h$-dependent CDW evolutions.

In this article, using extensive first-principles calculations (see Methods), we propose that the intrinsic EI and MEPC may be possible reasons for the CDW formation in $R$NiO$_2$. Remarkably, different from the low-energy single-orbital physics in CaCuO$_2$, the Ni 3 $d_{x2-y2}$ and $R$ 5$d_{z2}$ play a larger role in contributing to EI and MEPC, respectively. In particular, the van Hove singularity (VHS) near the Fermi level ($E_F$) can induce strong Fermi surface nesting (FSN) at a series of $k_z$ planes in undoped $R$NiO$_2$. As the $n_h$ increases, the reduced FSN of Ni 3$d_{x2-y2}$ due to the shift of VHS and the decreased occupation of $R$ 5$d_{z2}$ can gradually weaken the EI and MEPC in $R$NiO$_2$, respectively, eventually suppressing the CDW formation. These $n_h$-dependent behaviors of EI and MEPC in $R$NiO$_2$ are significantly different from those in CaCuO$_2$, due to the unique features of VHS and multiorbitals around $E_F$ in $R$NiO$_2$. Surprisingly, both EI and MEPC are insensitive to the temperature, which may result in a rather high temperature CDW phase in undoped $R$NiO$_2$ (at least) up to 400 K, as confirmed by our preliminary experimental measurements (see Methods).

***Multiorbital feature in $R$NiO$_2$.*** Figure 1(a) shows the schematic band structure of $R$NiO$_2$ and its analog CaCuO$_2$ in their nonmagnetic (NM) phase, extracted from the DFT calculations (Fig. S1 in Supplementary Information). In both $R$NiO$_2$ and CaCuO$_2$, the symmetry-allowed orbital coupling between in-plane $R$/Cu 3$d_{x2-y2}$ (majority) and O 2$p_x$+$p_y$ (minority) orbitals results in a delocalized band across the $E_F$. The bandwidth of 3$d_{x2-y2}$ in $R$NiO$_2$ is ~1 eV narrower than that in CaCuO$_2$ [24,25]. One of the significant differences between $R$NiO$_2$ and CaCuO$_2$ is the presence of partially occupied low-lying $R$ 5$d_{z2}$ state across $E_F$ in $k_z$=0 plane, which slightly hybridizes with Ni 3$d_{z2}$ orbital, leading to a multiorbital feature in $R$NiO$_2$ [26-29]. In contrast, Ca 3$d_{z2}$ is fully empty and located at ~2 eV above $E_F$, resulting in a single-orbital feature in CaCuO$_2$. Interestingly, the Lieb-type lattice formed by NiO$_2$ or CuO$_2$ plane creates VHSs at the boundary (X and R points) of Brillouin zone (BZ), as labeled by black arrows in Fig. 1(a). Importantly, in $R$NiO$_2$, the VHS is below $E_F$ in $k_z$=0 plane but slightly above $E_F$ in $k_z$=0.5$c^*$ plane, indicating that its $E_F$ will exactly cross VHS in a specific $k_z$ plane. However, differing from $R$NiO$_2$, all the VHSs in CaCuO$_2$ are located below $E_F$, partially due to the absence of Ca $5d_{z^2}$ state around $E_F$ induced over occupation of $3d_{x^2-y^2}$ in CaCuO$_2$. As illustrated below, the different locations of VHS between $R$NiO$_2$ and CaCuO$_2$ will lead to a delicate difference in their FS of $3d_{x^2-y^2}$ band and corresponding EI.

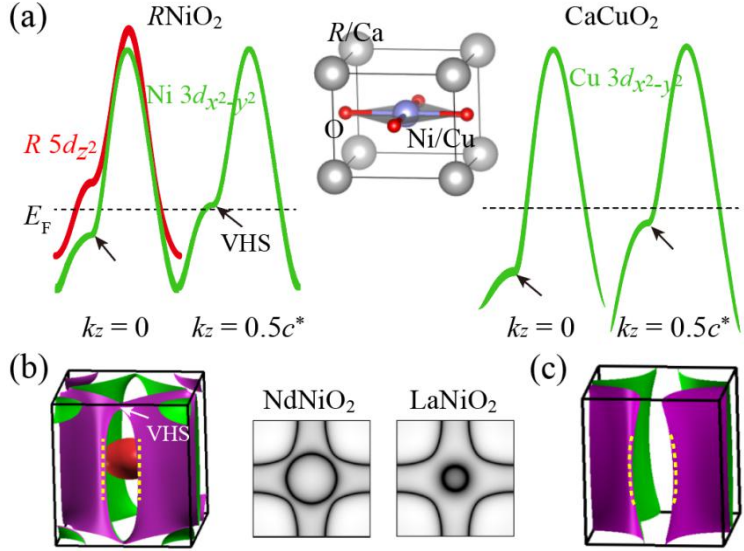

FIG. 1. (a) Comparison of low-energy electronic structures between $R$NiO$_2$ ($R$=Nd, La) and CaCuO$_2$ around $E_F$, in which only the dominated $d$ orbital characters are presented. Inset: structure of $R$NiO$_2$/CaCuO$_2$. (b) Left: 3D FS of NdNiO$_2$. Right: Cross sections of FSs in $k_z$=0 plane for NdNiO$_2$ and LaNiO$_2$. VHSs are labeled by arrows in (a) and (b). (c) 3D FS of CaCuO$_2$. Yellow-dashed lines in (b) and (c) highlight the FS evolutions along $k_z$.

The calculated FSs of undoped NdNiO$_2$ and CaCuO$_2$ are shown in Figs. 1(b) and 1(c), respectively. In both systems, the $3d_{x^2-y^2}$ orbital shows a cylinder-like hole pocket, exhibiting a quasi-2D-like dispersion. However, the existence of VHS at FS in NdNiO$_2$ but not in CaCuO$_2$ results in a noticeably different $3d_{x^2-y^2}$ FS between NdNiO$_2$ and CaCuO$_2$. When $k_z$ changes from 0.5$c^*$ to 0, near VHS [labeled by the arrow in Fig. 1(b)], there is a significant momentum variation of FS in $k_x$-$k_y$ plane; after passing through VHS, the energy band has a very large dispersion, and the shape of FS is almost unchanged in $k_x$-$k_y$ plane with small $k_z$, exhibiting a nearly parallel segment of FS [highlighted by parallel lines in Fig. 1(b)]. This unusual FS feature, which is absent in (undoped) CaCuO$_2$ [highlighted by curved lines in Fig. 1(c)], is critical for the formation of strong FSN in $R$NiO$_2$ (also see Supplementary Fig. S1 for FS in $k_y$ = 0.5$b^*$ plane). Another striking difference between NdNiO$_2$ and CaCuO$_2$ is the appearance of $5d_{z^2}$ state in the vicinity of $\Gamma$ point [Fig. 1(a)], which forms a sphere-like electron pocket in NdNiO$_2$ [Fig. 1(b)], exhibiting an isotropic 3D dispersion. As for NdNiO$_2$ and LaNiO$_2$, since the orbital energy of La $5d$ is higher than that of Nd $5d$, fewer $5d_{z^2}$ orbital is occupied in LaNiO$_2$ than in NdNiO$_2$, resulting in a smaller electron pocket in LaNiO$_2$ [Fig. 1(b)]. As discussed later, this difference will result in weaker EI and MEPC in

LaNiO$_2$ than in NdNiO$_2$. We emphasize that all these unique features of FS remain unchanged even under the hybrid functionals or dynamical mean-field theory calculations [30,31].

***Hole-doping-dependent EI and MEPC in RNiO$_2$.*** While the ground state of CaCuO$_2$ is an antiferromagnetic (AFM) Mott insulator [27], only local AFM fluctuations are observed in $R$NiO$_2$ [32,33], which even cannot coexist with the CDW phases in NdNiO$_2$ [12,13], possibly due to the competing between these different symmetry-breaking orders [3,4]. Meanwhile, although moderate hole doping can introduce superconductivity in both CaCuO$_2$ and $R$NiO$_2$, it could stabilize the CDW phases in CaCuO$_2$ [4] but weakens or even destroys the CDW phases in $R$NiO$_2$ [11,12]. Therefore, it is expected that the different low-energy orbital feature and FS (Fig. 1) could be responsible for different interplay between hole doping and CDW in $R$NiO$_2$ and CaCuO$_2$.

The real (Re$\chi$) and imaginary (Im$\chi$) parts of electron susceptibility (see Methods) reflect the EI and FSN of a system, respectively. Meanwhile, MEPC can be evaluated by calculating phonon linewidth $\gamma$ (see Methods). Generally, the Re$\chi(\boldsymbol{q})$ and/or $\gamma(\boldsymbol{q})$ diverging at a wave vector $\boldsymbol{q}$ might trigger a $\boldsymbol{q}$-modulated CDW [23]. In the following, we focus on the discussion of NdNiO$_2$ and briefly mention the results of LaNiO$_2$ in the main text.

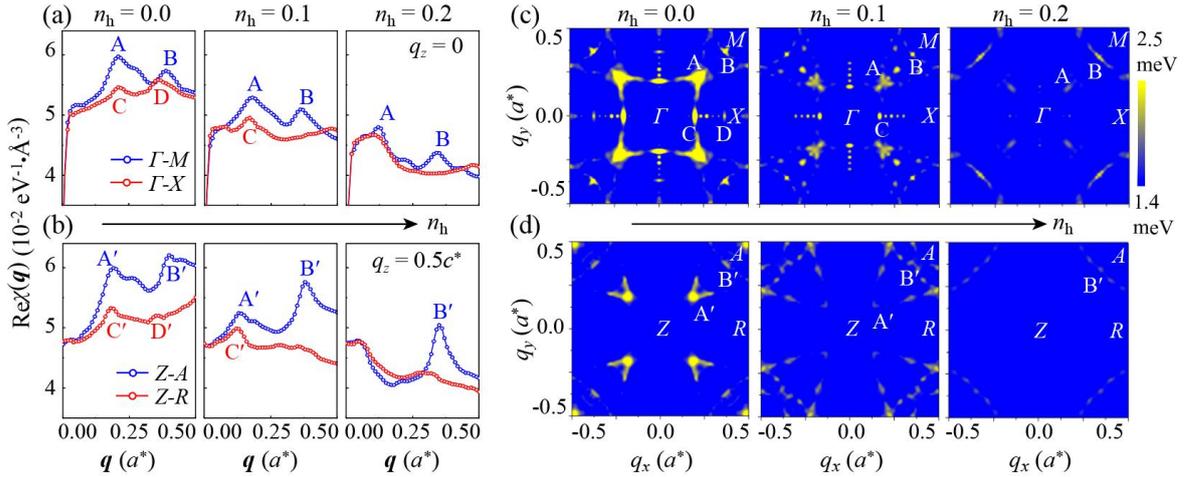

FIG. 2. (a) Calculated Re$\chi(\boldsymbol{q})$ of NdNiO$_2$ under three different $n_h$ (in units of hole/u.c.) along two high-symmetry lines in $q_z = 0$ plane. (b) Same as (a) but in $q_z = 0.5c^*$ plane. (c) Calculated $\boldsymbol{q}$-resolved $\gamma(\boldsymbol{q})$ in $q_z = 0$ plane from all phonon modes for electronic states in entire BZ. (d) same as (c) but in $q_z = 0.5c^*$ plane. A (A'), B (B'), C (C'), and D (D') indicate the corresponding peak positions in Re$\chi(\boldsymbol{q})$ and $\gamma(\boldsymbol{q})$ at different $\boldsymbol{q}$.

Figures 2(a) and 2(b) show the calculated $\boldsymbol{q}$-dependent Re$\chi(\boldsymbol{q})$ in NdNiO$_2$ along two high-symmetry lines in $q_z = 0$ and $q_z = 0.5c^*$ planes, respectively. Especially, the calculated Re$\chi(\boldsymbol{q})$ in the entire BZ (Supplementary Fig. S2) confirm that the maxima of Re$\chi(\boldsymbol{q})$ are located in these lines. Overall, when $n_h = 0$, there are some similarities in these two planes, *i.e.*, there are multiple high-intensity peaks along $\Gamma$-$M$ ($\Gamma$-$X$) and $Z$-$A$ ($Z$-$R$) appearing at similar $\boldsymbol{q}$, reflecting a quasi-2D character of band structures. Importantly, these corresponding peaks in Re$\chi(\boldsymbol{q})$ are also observed in Im$\chi(\boldsymbol{q})$ (Supplementary Fig. S3), indicating that FSN plays important roles in forming the peaks in Re$\chi(\boldsymbol{q})$. Surprisingly, as the increase of $n_h$, the peak intensities of A, B (B') gradually decrease, while the peaks of A', C (C'), and D (D') will eventually disappear; meanwhile, the corresponding $q$ for these peaks will gradually shift to smaller values. These unusual $n_h$-dependent behaviors are also observed in LaNiO$_2$ (Supplementary Fig. S4). To partially treat the electron-electron correlation effects, the on-site Hubbard $U$ effects are also considered, which can further strengthen our conclusions (Supplementary Figs. S5-S7). Opposing to $R$NiO$_2$, the peaks in Re$\chi(q)$ are very weak (almost absent) in undoped CaCuO$_2$ but become noticeable under certain $n_h$, along with the increased values of Re$\chi(q)$ (Supplementary Fig. S8).

Figures 2(c) and 2(d) show the calculated $q$-dependent $\gamma(q)$ in $q_z = 0$ and $q_z = 0.5c^*$, respectively. Overall, when $n_h = 0$, the $q$ for generating high-intensity peaks in $\gamma(q)$ are consistent with that in Re$\chi(q)$, although the relative intensities for different peaks are slightly different in Re$\chi(q)$ and $\gamma(q)$. Interestingly, the C' and D' peaks can only exist in Re$\chi(q)$ but not in $\gamma(q)$. Importantly, these peak values are comparable or even stronger than those critical $\gamma(q)$ values in MEPC-induced CDW systems, e.g., 1$T$-TaSe$_2$ [34], 1$T$-VSe$_2$ [35], and TbTe$_3$ [36]. When $n_h$ increases, the intensities of these peaks dramatically decrease, and some of them (e.g., A' and C) can even disappear at a high $n_h$. Therefore, it can be expected that, when $n_h = 0$, the joint contributions of multiple peaks in both Re$\chi(q)$ and $\gamma(q)$ from multiple $q_z$ planes might play important roles in generating 3D CDW (rather than 2D-like CDWs as commonly observed in cuprates [5]) in NdNiO$_2$. When $n_h$ increases, both EI and MEPC are strongly weakened, consequently suppressing the driving forces for CDW formation. Generally, a similar conclusion is found in LaNiO$_2$ (Supplementary Fig. S4) and, again, the consideration of on-site Hubbard $U$ can further strengthen our conclusion (Supplementary Figs. S6-S7). Opposing to the $R$NiO$_2$, the $\gamma(q)$ of CaCuO$_2$ are much insensitive to $n_h$ (Supplementary Fig. S8).

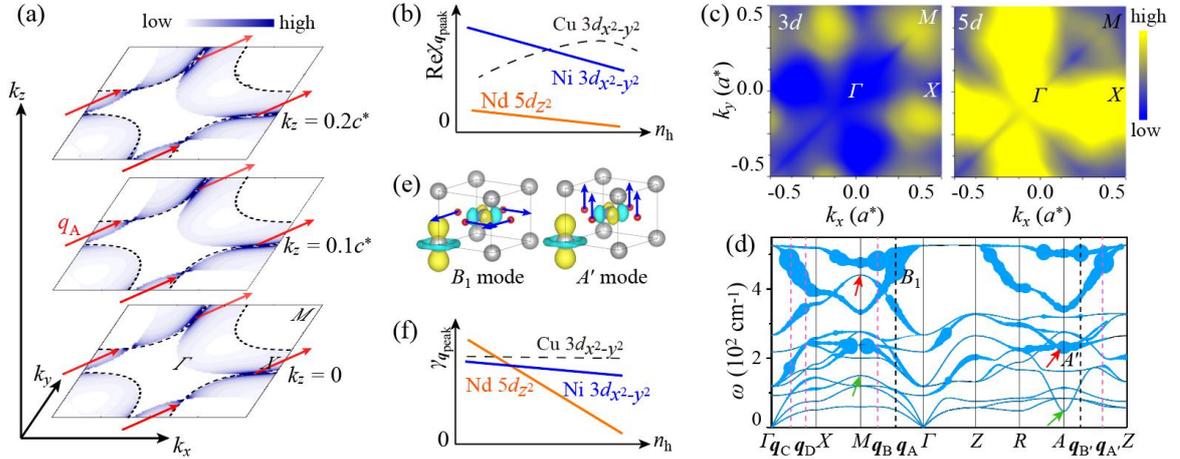

FIG. 3 (a) Evolution of $k$-resolved Re$\chi_q(k)$ at $q_A$ along different (small) $k_z$ in undoped NdNiO$_2$, where Re$\chi_q(k)$ is projected to Ni 3$d_{x2-y2}$ orbital. Black-dashed lines represent the FS of Ni 3$d_{x2-y2}$ band and red arrows of $q_A$ connect the electronic states on FS to form a nesting (in view of almost unchanged FS along $k_z$). (b) Contributions of Ni 3$d_{x2-y2}$ and Nd 5$d_{z2}$ to Re$\chi_{qpeak}$ as a function of $n_h$. (c) MEPC matrix $g(k)$ from all the phonon modes at $q_A$ in $k_z = 0$ plane. (d) Phonon spectrum of undoped NdNiO$_2$ with the magnitudes of phonon linewidth indicated by the line thickness. (e) Two typical strong orbital-phonon coupling modes at $q_A$ and $q_{B'}$, labeled in (d). (f) Contributions of Ni 3$d_{x2-y2}$ and Nd 5$d_{z2}$ to $\gamma_{qpeak}$ as a function of $n_h$. Contributions of Cu 3$d_{x2-y2}$ to Re$\chi_{qpeak}$ and $\gamma_{qpeak}$ of $CaCuO_2$ as a function of $n_h$ are also schematically plotted in (b) and (f), respectively, for comparison with the Ni $3d_{x2-y2}$ and Nd $5d_{z2}$ of $NdNiO_2$.

***Mechanism for hole-doping-dependent EI and MEPC.*** To further investigate the contribution of electronic states to $Re\chi(q)$ in $NdNiO_2$, taking peak A as an example (other peaks are similar), the ***k***-resolved $Re\chi_{qA}(k)$ for Ni $3d_{x2-y2}$ band (a majority contributor) is shown in Fig. 3(a), and the results for Nd $5d_{z2}$ (a minority contributor) are given in Supplementary Information (Supplementary Figs. S9 and S10). Basically, for a fixed ***q***, the contributions to $Re\chi$ can be divided into two parts: (i) states on FS connected by ***q*** [dark-blue spots in Fig. 3(a)]; (ii) occupied and unoccupied states (not on FS) connected by ***q*** [light-blue regions in Fig. 3(a)] (see Supplementary Fig. S9 for details). Importantly, while (i) strongly depends on the shape of FS and determines the peak formation to $Re\chi$, (ii) is weak and mostly contributes a uniform value to $Re\chi$. As shown in Fig. 3(a), the unique FS shape of $NdNiO_2$, induced by the VHS across $E_F$, makes the states on FS near BZ boundary well connected by ***q***$_A$ [red arrows in Fig. 3(a)], contributing a large value to $Re\chi_{qA}$. Importantly, the FS in $k_x$-$k_y$ plane is almost unchanged with variable $k_z$ between $0 \le k_z \le 0.2c^*$ [*e.g.*, parallel lines marked in Fig. 1(b)], and the connection vector ***q***$_A$ in different $k_z$ planes keeps constant in this FS segment (also see Supplementary Fig. S10 for details), resulting in a strong FSN and contributing a large peak A to $Re\chi(q)$ [Fig. 2(a)]. As a comparison, the FS of (undoped) $CaCuO_2$ has a variation with $k_z$ due to the absent VHS across $E_F$, and the connection vector ***q*** is also changed gradually (Supplementary Fig. S10). Consequently, there is no obvious peak in $Re\chi(q)$ of undoped $CaCuO_2$ (Supplementary Figs. S8 and S10).

Under hole doping, the VHS across $E_F$ shifts to a smaller $k_z$ value, which reduces the $k_z$ range for parallel FS segment of $3d_{x2-y2}$ band [*i.e.*, the length of parallel lines marked in Fig. 1(b)], thus decreasing the strength of FSN (Supplementary Fig. S11). In addition, hole doping reduces the occupation of Nd $5d_{z2}$ states at FS, decreasing their sphere-like electron pocket. Fig. 3(b) shows the change of contribution from Ni $3d_{x2-y2}$ and Nd $5d_{z2}$ states to $Re\chi_{qA}$ (and other peaks) as a function of $n_h$ (also see Supplementary Fig. S10). As $n_h$ increases, the contribution of both Ni $3d_{x2-y2}$ and Nd $5d_{z2}$ decrease, explaining the observation in Fig. 2(a). Meanwhile, as $n_h$ increases, the evolved FS gradually shifts ***q***$_A$ to smaller values. On the other hand, the opposite trend of $Re\chi_q$ in $CaCuO_2$ under hole doping can be well understood as the gradual approach of VHS to $E_F$ and the lack of $5d$ orbital (Supplementary Fig. S11), *i.e.*, a dome region could exist in $CaCuO_2$ but not in $RNiO_2$.

In contrast to $Re\chi_q(k)$, Nd $5d_{z2}$ plays a larger role in MEPC. As shown in Fig. 3(c), the MEPC matrix $g(k)$ at ***q***$_A$ obtained from all the phonon modes highlight the major contribution of intraband scattering of Nd $5d_{z2}$. The magnitude of interband scattering between Ni $3d_{x2-y2}$ and Nd $5d_{z2}$ is similar to the intraband scattering of Ni $3d_{x2-y2}$ orbital (Supplementary Figs. S12 and S13). The calculated phonon spectrum of undoped $NdNiO_2$ is shown in Fig. 3(d). Comparing the phonon dispersion between $k_z = 0$ and $k_z = 0.5c^*$ planes, it is clearly identified that there are two strongly soft optical modes around $A$, as indicated by the arrows. The projected $\gamma(q)$ shows that the main contribution comes from optical modes. Importantly, one soft mode partially contributes to the large $\gamma_{qB'}$. The phonon spectrum of $LaNiO_2$ along with the projected $\gamma(q)$ are similar to that of

NdNiO$_2$ (Supplementary Fig. S14). Importantly, the appearance of soft optical modes is also observed in CDW-phase LaNiO$_2$ in experiments [11].

Figure 3(e) shows the two typical strong orbital-phonon coupling (OPC) modes at $q_A$ and $q_{B'}$, as marked in Fig. 3(d). Similar to that in cuprates [37], due to the large mass differences between Nd/Ni and O atoms, the atomic vibrations are much strong for O atoms. While the orbital components in both modes come from the Nd $5d_{z2}$ and Ni $3d_{x2-y2}$, the phonon modes can be either in-plane $B_1$ mode or out-of-plane $A'$ mode. The $B_1$ mode contribute large EPC, and the $A'$ mode relates to the soft phonon. When $n_h$ increases, the phonon spectrum of NdNiO$_2$ only have small changes (Supplementary Fig. S14), but the occupation of Nd $5d_{z2}$ around $E_F$ dramatically decreases. Consequently, as shown in Fig. 3(f), the $\gamma_{qA}$ decreases rapidly, explaining the observation in Figs. 2(c) and (d). On the other hand, the lack of Ca $5d_{z2}$ around $E_F$ can explain the much insensitivity of $\gamma_q$ under different $n_h$ in CaCuO$_2$ (Supplementary Fig. S8).

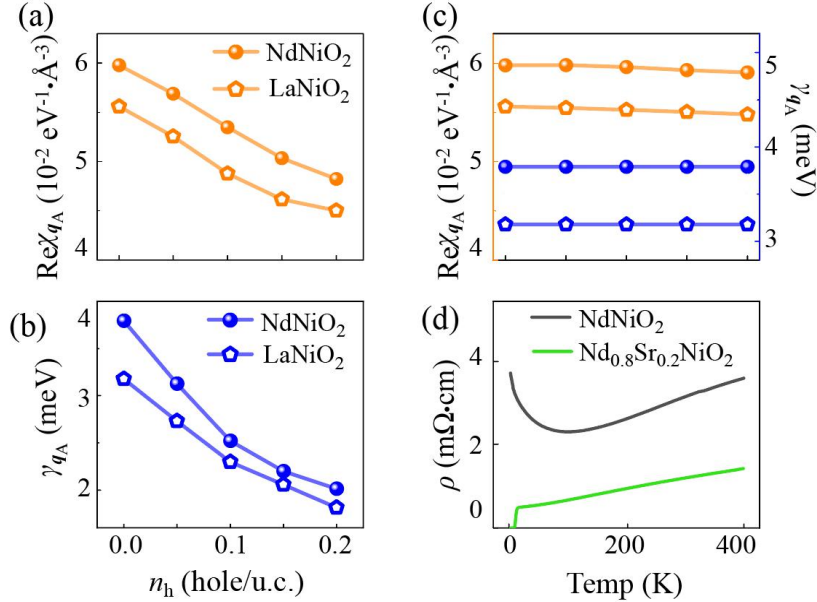

FIG. 4 Comparison of calculated (a) Re$\chi_{qA}$ and (b) $\gamma_{qA}$ for NdNiO$_2$ and LaNiO$_2$ as a function of $n_h$. (c) Calculated Re$\chi_{qA}$ and $\gamma_{qA}$ as a function of temperature. (d) Experimentally measured electrical resistivity of NdNiO$_2$ (with CDW) and Nd$_{0.8}$Sr$_{0.2}$NiO$_2$ (without CDW) as a function of temperature.

***Comparison between NdNiO$_2$ and LaNiO$_2$.*** It is interesting to compare the calculated Re$\chi(q)$ and $\gamma(q)$ between NdNiO$_2$ and LaNiO$_2$. Overall, the spectra of Re$\chi(q)$ or $\gamma(q)$ are similar for both systems, except for the specific values. As shown in Figs. 4(a) and 4(b), the values of Re$\chi_{qA}$ and $\gamma_{qA}$ in LaNiO$_2$ are smaller than that of NdNiO$_2$, mostly because the occupation of La $5d_{z2}$ is smaller than that of Nd $5d_{z2}$ [Fig. 1(b)]. Under hole doping, obeying the mechanism discussed above, the values of Re$\chi_{qA}$ and $\gamma_{qA}$ will gradually decrease in both systems, accompanied by the decreases of $q_A$. Generally, the trend of other peaks in Re$\chi_q$ and $\gamma_q$ and their $n_h$-dependent behaviors (Supplementary Fig. S15) are similar to that of Re$\chi_{qA}$.

Unexpectedly, as shown in Fig. 4(c), the values of Re$\chi_{q\text{A}}$ and $\gamma_{q\text{A}}$ are estimated to be insensitive to the electron temperature even up to 400 K for the undoped $R$NiO$_2$ (similar for other $q$ values, Supplementary Fig. S16), indicating the strong EI and MEPC may maintain above room temperature. Therefore, if the CDWs observed in $R$NiO$_2$ partially originate from the EI and MEPC, it may also survive at high temperature. In experiments, the undoped NdNiO$_2$ grown on SrTiO$_3$ substrate (without STO capping layers) may exhibit a CDW phase [12]. To confirm our theory, we perform the transport measurements using a similar CDW-phase NdNiO$_2$ sample adopted in Ref. 12 (see Methods). As shown in Fig. 4(d), no signal of a sharp phase transition is observed up to 400 K in the electrical resistivity, indicating that the CDW phase in NdNiO$_2$ might survive at such a high temperature. For our Nd$_{0.8}$Sr$_{0.2}$NiO$_2$ sample with a superconducting temperature of ~10 K, the CDW phase is not observed to coexist with the superconductivity using RIXS measurement [12], possibly due to their competitive relationship and/or largely reduced EI and MEPC in heavy-hole-doped $R$NiO$_2$.

***Discussion and Outlook.*** It is important to compare our theory with the existing experimental observations. First, it is observed in experiments that the $R$ 5$d_{z2}$ and Ni 3$d_{x2-y2}$ orbitals are resonant during the formation of CDWs in $R$NiO$_2$ [11,12], consistent with our theoretical calculations. More importantly, beyond the experimental observations, our theory further reveals that the Ni 3$d_{x2-y2}$ [Fig. 3(a)] and $R$ 5$d_{z2}$ [Fig. 3(c)] play the major role in forming strong EI and MEPC in undoped $R$NiO$_2$, respectively, which may in turn result in a very high temperature CDW phase (at least) up 400 K, as confirmed by our preliminary experimental measurements. In comparison, the CDWs observed in cuprates usually cannot exist above ~250 K [4,5].

Second, it is observed in experiments that hole doping gradually destabilizes or even suppresses the CDW phase in $R$NiO$_2$ [11,12], which is strongly different from (even opposite to) that in cuprates [5,14]. It may also be explained by our theory. Especially, our theory demonstrates that as $n_\text{h}$ increases, the shift of VHS and the rapidly decrease of $R$ 5$d_{z2}$ orbital at FS largely weaken EI [Fig. 3(b)] and MEPC [Fig. 3(f)] in $R$NiO$_2$, respectively, suppressing the driving forces for CDW formation in $R$NiO$_2$. We note that the part role of strong correlation effect in CDW formation may not be excluded, even there is no direct evidence that it is a significant change of correlation by hole doping [20,21] that suppress the CDWs. Again, the recent calculations from both hybrid functional and dynamical mean-field theory [30,31] present almost the same low-energy electronic structure and FS as ours.

Finally, it is still a challenge to directly identify the $q_\text{CDW}$ based on our calculations, as it likely be a combination of several similar low-energy CDW excitations at multiple peaks in multiple $q_z$ planes of Re$\chi_q$ and $\gamma_q$. These multiple peaks also indicate that multiple $q_\text{CDW}$ might appear under different conditions, calling for future experimental verification. However, our theory predicts that the $q_\text{CDW}$ in LaNdO$_2$ and NdNiO$_2$ could be very similar due to their nearly identical peaks in Re$\chi(q)$ and $\gamma(q)$, consistent with the experimental observations [11,12]. In practice, the exact $q_\text{CDW}$ could also highly depends on the local environments, defects (*e.g.*, possible H impurity [13,38]), and structural disorders [11,39], which are not included in the present study.

***Acknowledgement.*** We thank Dr. J. G. Guo at IOP for the help on the measurements of electrical resistivity of $NdNiO_2$ and $Nd_{0.8}Sr_{0.2}NiO_2$ at different temperatures and Dr. Z. Liu at Tsinghua University for the valuable comments on our results. We acknowledge the support from the NSFC (12004030, 12088101) and the NSAF (U1930402). The calculations were performed at Tianhe2-JK at CSRC.

# Methods

## 1. Computational methods

All the electronic structures and crystal structure optimizations are performed using the first-principles calculations as implemented in the Quantum ESPRESSO (QE) [40], and the scalar relativistic ultrasoft pseudopotentials are used [41]. The kinetic energy cutoff for plane waves is set to 80 Ry. Some calculations are also tested using the Vienna ab initio simulation package (VASP) [42] within the projector augmented wave method [43] and Perdew-Burke-Ernzerhof [44] exchange correlation functional, which give the same results as QE. The lattice constants in $xy$ plane are fixed at $a = b = 3.905$ Å to match the SrTiO$_3$ substrate, while lattice in $z$ direction is fully relaxed, which gives $c = 3.34$ Å and 3.45 Å for NdNiO$_2$ and LaNiO$_2$, respectively. The van der Waals correction is adopted [45]. The electron-phonon coupling calculations are performed by Wannier interpolation [46,47], as implemented in EPW code [48], which allows us to achieve sufficiently accurate $k$-point sampling in the BZ. The calculations of the phonon linewidth are performed on dense grids of (100×100×100) $k$-points and (100×100×1) $q$-points. The on-site Hubburd $U$ [49,50] is also considered, which is added on the $3d$ orbitals of Ni. The temperature effect is considered as implemented in the Fermi-Dirac distribution.

The real Re$\chi(q)$ and imaginary Im$\chi(q)$ part of electron susceptibility [51] are calculated by

$$\mathrm{Re}\chi(q) = \sum_k \frac{f(\varepsilon_k) - f(\varepsilon_{k+q})}{\varepsilon_k - \varepsilon_{k+q}} \quad (1)$$

$$\mathrm{Im}\chi(q) = \sum_k \delta(\varepsilon_k - \varepsilon_F)\delta(\varepsilon_{k+q} - \varepsilon_F),$$

where $f(\varepsilon_k)$ is the function of Fermi-Dirac distribution. The real part reflects the stability of the electronic system, and the imaginary part reflects the Fermi surface topology of a system. The phonon linewidth $\gamma(q)$ which directly reflects the electron-phonon coupling strength is defined as

$$\gamma(q,\nu) = 2\pi\omega_{q\nu} \sum_{mn} \int \frac{dk}{\Omega_{BZ}} |g_{mn,q\nu}(k)|^2 \times \delta(\varepsilon_{m,k} - \varepsilon_F)\delta(\varepsilon_{n,k+q} - \varepsilon_F), \quad (2)$$

where $m$, $n$ are the band indexes, and $\nu$ represents the phonon mode. The coupling matrix is

$$g_{mn,q\nu}(k) = \frac{1}{\sqrt{2\omega_{q\nu}}} \langle \varphi_{m,k} | \partial_{q\nu} V | \varphi_{n,k+q} \rangle. \quad (3)$$

The $\varphi_{m,k}$ is the electronic wave function, with eigenvalue $\varepsilon_{m,k}$.

## 2. Experimental methods

The NdNiO$_2$ films with infinite-layer structure are prepared by topochemical reduction of perovskite NdNiO$_3$ without capping layer. NdNiO$_3$ films (thickness about 10 nm) are deposited on TiO$_2$-terminated STO (001) substrates by 248-nm KrF laser. During the deposition, the substrate temperature is controlled at 620 °C with the oxygen pressure of 200 mTorr. A laser fluence of 1.2 J/cm$^2$ is used to ablate the target and the size of laser spot is about 3 mm$^2$. After deposition, the samples are cooled down in the same oxygen pressure at the rate of 10 °C/min. In order to acquire the infinite-layer nickelate phase, the as-grown samples are sealed in the quartz tube together with

0.1g $CaH_2$. The pressure of the tube is about 0.3 mTorr. Then, the tube is heat up to 290 °C in tube furnace and hold for 2 hours and cooled down naturally with the ramp rate of 10 °C/min. The superconducting $Nd_{0.8}Sr_{0.2}NiO_2$ films are prepared in the same way as the parent $NdNiO_2$. The crystal structures of films are characterized using a Bruker D8 Discover diffractometer. The temperature-dependent resistivity is measured using four-probe method in a physical properties measurement system (PPMS, Quantum Design Inc.).